\documentclass[amssymb,nobibnotes,superscriptaddress,longbibliography,notitlepage,twocolumn]{revtex4-1}

\usepackage{enumerate,amsmath}
\usepackage{graphicx}
\usepackage{color} 

\usepackage{braket}

\newtheorem{theo}{Theorem}
\newtheorem{defi}{Definition}

\begin{document}

\title{Quantum speedup in stoquastic adiabatic quantum computation}%

\author{Keisuke Fujii}
\address{Department of Physics, Graduate School of Science, Kyoto University, Kitashirakawa-Oiwakecho, Sakyo-ku, Kyoto 606-8502, Japan}
\address{JST, PRESTO, 4-1-8 Honcho, Kawaguchi, Saitama, 332-0012, Japan}

\date{\today}
\begin{abstract}
Quantum computation provides exponential speedup 
for solving certain mathematical problems against classical computers.
Motivated by current rapid experimental progress on 
quantum computing devices, various models of quantum computation 
have been investigated to show quantum computational supremacy.
At a commercial side, 
quantum annealing machine realizes the quantum Ising model 
with a transverse field and heuristically 
solves combinatorial optimization problems. 
The computational power of this machine 
is closely related to adiabatic quantum computation (AQC) 
with a restricted type of Hamiltonians, namely stoquastic Hamiltonians,
and has been thought to be relatively less powerful 
compared to universal quantum computers.
Little is known about computational quantum speedup 
nor advantage
in AQC with stoquastic Hamiltonians.
Here we characterize computational capability of 
AQC with stoquastic Hamiltonians, which we call stoqAQC.
We construct a concrete stoqAQC model,
whose lowest energy gap is lower bounded polynomially, and hence 
the final state can be obtained in polynomial time.
Then we show that it can simulate universal quantum computation 
if adaptive single-qubit measurements
in non-standard bases are allowed on the final state.
Even if the measurements are restricted to non-adaptive measurements
to respect the robustness of AQC,
the proposed model exhibits quantum computational supremacy;
classical simulation is impossible under complexity theoretical conjectures.
Moreover, it is found that such a stoqAQC model can simulate 
Shor's algorithm and solve the factoring problem in polynomial time.
We also propose 
how to overcome the measurement imperfections via quantum error correction
within the stoqAQC model and also
an experimentally feasible verification scheme 
to test whether or not stoqAQC is done faithfully.
\end{abstract}


\maketitle
\section{Introduction}
The Hamiltonians with
nonpositive off-diagonal elements 
in a standard basis
are called stoquastic Hamiltonians~\cite{stoqBravyi}.
Important quantum statistical models are 
included in this class. 
The quantum Ising model with a transverse field,
the antiferromagnetic Heisenberg model on bipartite graphs, 
and the Bose-Hubbard model with negative hoppings
are such examples.
For this class of models,
the ground state
has real and positive coefficients 
in the standard basis.
This allows 
a Monte-Carlo method on a classical computer,
called quantum Monte-Carlo,
to sample the ground state,
while efficient convergence
is not always guaranteed.

Another important aspect of stoquastic Hamiltonians 
is quantum annealing (QA)~\cite{QA}, which 
is a heuristic algorithm to 
solve combinatorial optimization problems 
approximately by adiabatically (or even non-adiabatically) 
changing the parameters.
In the adiabatic case,
QA is included in an adiabatic quantum  computation (AQC)~\cite{AQC},
which is known to be universal 
for quantum computation
when nonstoquastic Hamiltonians are employed~\cite{AharonovAQC}.
When it is restricted to stoquastic Hamiltonians,
its computational power has been speculated to be less powerful 
compared to universal quantum computation~\cite{AQCreview,NISQ}.
From an experimental viewpoint, 
stoquastic Hamiltonians are relatively 
easy to be implemented.
This is why QA with the quantum Ising model
with a transverse field has been already 
implemented with a larger number of qubits 
on a superconducting system~\cite{Dwave11,Dwave14,DwaveNatPhys},
while the quality of the qubits are relatively poor 
compared to the standard circuit-based approaches~\cite{Google14,Google15}.
Unfortunately, there has been little theoretical or experimental 
solid evidence of quantum speedup 
in AQC with stoquastic Hamiltonians~\cite{AQCreview,NISQ}
except for the oracle problems~\cite{GroverAQC1,GroverAQC2,Nagaj}.

Computational complexity of stoquastic Hamiltonians 
has been investigated so far
in various aspects~\cite{stoqBravyi,stoqBravyi06,stoqBravyi17}.
Local Hamiltonian problems of stoquastic Hamiltonians
have been shown to be stoqMA-complete~\cite{stoqBravyi06}.
Moreover, the local Hamiltonian problem of the quantum Ising model with 
a transverse field on degree-3 graphs is 
complete in the sense that 
it is equivalent modulo polynomial reductions 
to the local Hamiltonian problems of stoquastic Hamiltonians~\cite{stoqBravyi06,stoqBravyi17}.
StoqMA is the class of problems 
efficiently verifiable by reversible (unitary) classical computation, 
with an $X$-basis measurement,
whose input consists of 
a quantum state provided by a prover as the proof  
and ancilla qubits prepared in the Pauli $X$ and $Z$ bases.
While stoqMA is not so powerful as 
QMA~\cite{KitaevBook,KempeQMA},
stoqMA includes MA and hence NP,
which are efficiently verifiable problems by 
probabilistic and deterministic classical computations,
respectively.
This implies that
computational complexity of the ground state energy of stoquastic 
Hamiltonians does not directly 
reflect computational power of 
AQC with stoquastic Hamiltonians;
the former is thought to be much harder than 
the latter.

The quench dynamics of stoquastic Hamiltonians 
is as powerful as universal quantum computation.
Quantum computational supremacy of quantum approximated 
optimization algorithm~\cite{QAOA}, which is a digitalized version of 
QA, has been argued~\cite{FrahiHarrow} by using the fact that 
Ising type commuting interactions on the eigenstates of 
the transverse field can simulate non-universal model,
so-called IQP~\cite{IQP0,IQP,FujiiMoriIQP} (instantaneous quantum polynomial time computation).
Even translation invariant quench dynamics on one-dimensional
quantum Ising model can simulate universal quantum computation~\cite{RaussendorfIsing}. 
Note that, quench dynamics of stoquastic Hamiltonians can 
generate negative (even complex) coefficients in the standard basis.
Moreover, quench dynamics does not inherit the robustness of 
AQC, i.e., protection as a ground state.
Regarding adiabatic dynamics,
the computational power of stoquastic Hamiltonians
has not yet fully understood.
Neither quantum computational supremacy nor the capability of 
universal quantum computation has been addressed so far.

Here we characterize computational capability of AQC 
with stoquastic Hamiltonians, which we call {\it stoqAQC}, 
and show that stoqAQC exhibits strong quantum speedup.
Based on AQC using the Feynman-Kitaev Hamiltonian~\cite{Feynman86,KitaevBook,AharonovAQC}, 
we construct a stoqAQC model whose 
lowest energy gap is always lower bounded by 
the inverse of a polynomial function in the size of the system.
This guarantees that the final state can be faithfully 
obtained with a polynomial time.
Specifically, 
we consider both non-adaptive and adaptive single-qubit
measurements in the Pauli bases on the final state of the stoqAQC.
While this contains a non-standard basis,
it is not so difficult to perform non-standard basis
measurements, if the actual quantum machine works coherently.

In the case of the adaptive single-qubit measurements,
we can successfully show that 
stoqAQC can simulate universal quantum computation
only Pauli basis measurements.
However, such adaptive single-qubit measurements 
take time for the sequential measurements, and 
the final state would decohere during the measurements.
Therefore, non-adaptive measurements might 
be relevant to characterize an actual stoqAQC machine.
Even in the case of the non-adaptive measurements, 
we can show that stoqAQC can perform non-universal quantum computation,
which exhibits quantum speedup.
More precisely, 
stoqAQC can simulate
two types of 
non-universal models for quantum computational 
supremacy, IQP~\cite{IQP0,IQP} and HC1Q (Hadamard-classical circuit with one-qubit)~\cite{HC1Q} models.
Moreover, we can show that stoqAQC 
can also simulate Simon's algorithm~\cite{Simon} 
and Shor's factorization algorithm efficiently~\cite{Shor,KitaevPhase}.
To this end, 
we slightly modify the phase estimation 
so that the phase is obtained with 
non-adaptive measurements without quantum Fourier transformation.
In this way, 
we find strong evidence that an ideal stoqAQC machine 
can provide plenty of quantum speedup.

Yet, this result does not mean that 
the state-of-the-art d-wave quantum annealer
designed to solve optimization problems
readily exhibits quantum speedup,
since it consists of relatively poor qubits and 
only employs the standard basis measurements.
In our construction, 
the final state is a highly entangled state
and the measurements are done in non-standard bases.
Therefore noise or imperfection in the system 
would affect the output crucially.
Moreover, the standard basis measurements only result in 
classical randomized computation in our construction.
Therefore, coherence in the final state is crucially important 
to gain quantum speedup in our construction,
though this would always be the case for any quantum computing device.

We further show that our construction is robust against 
measurement imperfections
by showing how to embed quantum error correction within the stoqAQC model.
We also argue how to verify whether or not the
proposed stoqAQC is faithfully done experimentally. 

\section{Stoquastic Adiabatic quantum computation}
We adopt the following definition of AQC:
\begin{defi}[AQC~\cite{AQCreview,AharonovAQC}]
\label{def1}
A $k$-local adiabatic quantum computation is specified by two $k$-local Hamiltonians $H_{\rm initial}$ and $H_{\rm final}$. 
The ground state of $H_{\rm initial}$ is unique and is a product state. 
The output is a state that is $\epsilon$-close in $l_2$-norm to the ground state of $H_{\rm final}$.
Let $s(t):[0,t_{f}] \mapsto [0,1]$ (the schedule) and 
let $t_f$ be the smallest time such that the final state of an adiabatic evolution generated by $H(s) = [1-s(t)]H_{\rm initial}+ H_{\rm final}$ 
for time $t_f$ is $\epsilon$-close in $l_2$-norm to the ground state of $H_{\rm final}$.
Then, arbitrary single-qubit measurements can be done 
adaptively or non-adaptively on the final state.
\end{defi}
\begin{defi}[StoqAQC~\cite{AQCreview}]
\label{def2}
Stoquastic adiabatic quantum computation (stoqAQC) is the special case of AQC  restricted to $k$-local stoquastic Hamiltonians.
\end{defi}
Following Ref.~\cite{AQCreview},
we here use the term QA
when stoquastic Hamiltonians are employed 
to solve combinatorial optimization problems
either adiabatically or non-adiabatically.
Our construction 
employs a stoquastic version of 
AQC using the Feynman-Kitaev Hamiltonian~\cite{Feynman86,KitaevBook,AharonovAQC}.
We consider a composite system 
$\mathcal{H}_{\rm work} \otimes \mathcal{H}_{\rm clock}$
of the working system $\mathcal{H}_{\rm work} = (\mathbb{C}^2)^{\otimes (n+m)}$
and the clock system $\mathcal{H}_{\rm clock}= \mathbb{C}^{T+1}$,
where $T={\rm poly}(n,m)$.
Later, the clock system is replaced by a $(T+1)$-qubit system 
by using the domain wall clock construction,
$|t\rangle = |1_{1}...1_{t} 0_{t+1}...\rangle$~\cite{KitaevBook,KempeQMA}.

The Hamiltonian is given by
\begin{eqnarray}
H(s)= (1-s) H_{\rm initial} + s H _{\rm final},
\end{eqnarray}
where
\begin{eqnarray}
H_{\rm initial} &=& 
H_{\rm in}
+ (I_c - |0\rangle \langle 0| _c ),
\\
H_{\rm final} &=& H_{\rm in}+ \sum _{t=1}^{T} \frac{1}{2} [ 
|t\rangle \langle t|_c + |t-1\rangle \langle t-1 |_c 
\nonumber\\ &&
- (U_t  |t\rangle \langle t-1| _c + {\rm h.c.} ) ],
\end{eqnarray}
with the energy penalty term for the initial state of the working system,
\begin{eqnarray}
H_{\rm in}=\sum _{i=1}^{n} |1\rangle \langle 1|_i \otimes |0\rangle \langle 0 | _c 
+\sum_{j=n+1}^{m+n} |-\rangle \langle - |_j  \otimes |0\rangle \langle 0 | _c ,
\end{eqnarray}
imposing the initial state $|0\rangle^{\otimes n} |+\rangle ^{\otimes m}$ .
The unitary operator $U_t$ 
acting on the working system
consists only of {\rm Toffoli}, {\rm CNOT}, and $X$.
This guarantees that $H(s)$ is stoquastic for $0\leq s \leq 1$
in the standard basis 
\begin{eqnarray}
\{ |0\rangle , |1\rangle \}^{\otimes (n+m)} \otimes \{ |t\rangle_c \}_{t=0}^{T} .
\end{eqnarray}

The parameter $s$ is adiabatically changed from $s=0$ to $s=1$.
The ground state of the initial Hamiltonian 
$H_{\rm initial}$ is 
\begin{eqnarray}
|\eta _{0} \rangle =  |0\rangle ^{\otimes n} |+\rangle ^{\otimes m} |0\rangle _c.
\end{eqnarray}
The ground state of the final Hamiltonian 
$H_{\rm final}$ is 
\begin{eqnarray}
|\Psi \rangle = \frac{1}{\sqrt{T+1}} \sum _{t=0}^{T} | \eta _t\rangle,
\end{eqnarray}
where 
\begin{eqnarray}
|\eta _t \rangle =  U_t \cdots U_1 | 0\rangle ^{\otimes n} |+\rangle ^{\otimes m} |t\rangle_c
\end{eqnarray}
Furthermore, 
regarding the minimum energy gap $\Delta$ 
between the ground and first excited state of $H(s)$ ($0\leq s \leq 1$),
we can make
the same argument as Ref.~\cite{AharonovAQC}
for AQC with general Hamiltonians.
This guarantees $\Delta$ is 
lower bounded by $O(1/T^{2})$.
By virtue of the adiabatic theorem,
there exist a certain constant $c(k)$
for any constant $k$ such that
if the computation time is sufficiently long
\begin{eqnarray}
t \geq c(k)\frac{ \| H_{\rm final} - H_{\rm initial} \|^{1+1/k}}{\epsilon ^{1/k} \Delta^{2+1/k}},
\label{eq:adiabatic}
\end{eqnarray} 
then the final state $|\Psi _{\rm ad}\rangle$ 
is $\epsilon$-close to the exact ground state $|\Psi \rangle$
in $l_2$-norm~\cite{ReichardtAQC,AharonovAQC,AQCreview}.

On the current QA machine,
measurements are done only in the standard computational basis,
in which the Hamiltonian is stoquastic.
Here we slightly relax this condition,
and assume that arbitrary single-qubit measurements
can be done on the final state either adaptively or non-adaptively.
(Note that the measurement basis is not 
restricted in Defs.~\ref{def1} and ~\ref{def2}
as is also stated in Ref.~\cite{AQCreview}.)
Suppose the clock state is measured, and 
the state $|T\rangle$ is obtained,
which occurs with a high probability $\geq 1/{\rm poly}(n,m)$.
It is known that this probability can be further amplified 
to a constant value~\cite{AharonovAQC}, say 1/2, 
by inserting the identity gates $(T-1)$ times
using the clock system $\{ | t \rangle_c\}_{t=0}^{2T-1}$.
The ground state is given by 
\begin{eqnarray}
|\Psi \rangle 
= \frac{1}{\sqrt{2T}} \sum _{t=0}^{2T-1} | \eta _t \rangle,
\end{eqnarray}
where $U_t = I$ for $T+1 \leq t \leq 2T-1$.
Any clock states in the rage $T \leq t \leq 2T-1$ allow
us the desired computation.

\begin{figure}
\center{\includegraphics[width=1\linewidth]{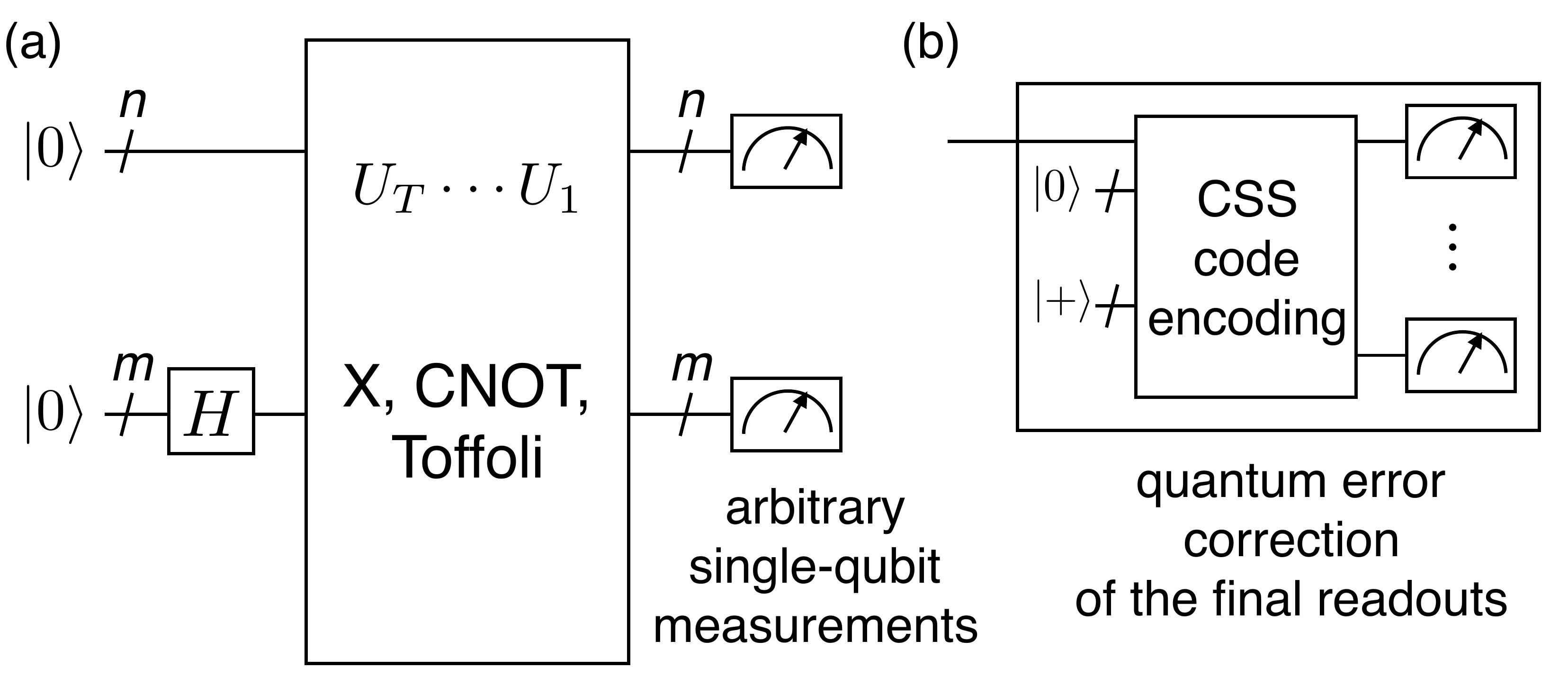}}%
\caption{(a) Quantum circuit that can be done by stoqAQC.
(b) Fault-tolerant readouts of the final state in the Pauli bases.
\label{fig1}}
\end{figure}

Under the condition of obtaining 
a successful clock state,
the computation, which we can perform on the final state,
is specified to be
\begin{eqnarray}
U_T \cdots U_1 (I^{\otimes n} \otimes H^{\otimes m})|0\rangle ^{\otimes (n+m)}
\end{eqnarray}
followed by arbitrary 
non-adaptive or adaptive single-qubit measurements
as shown in Fig.~\ref{fig1} (a).
If the measurements in Fig.~\ref{fig1} are restricted to the computational basis,
it corresponds to classical randomized computation.
When measurements on $xy$-plane are additionally available,
it belongs to Fourier hierarchy (FH) of the second level, FH${}_2$~\cite{FH,VerificationFH}.
If the initial states and measurements are restricted into 
$|0\rangle ^{\otimes n} |+\rangle$ and 
$X$ and $Z$ basis measurements on them
, the circuit in Fig.~\ref{fig1}
corresponds to HC1Q model~\cite{HC1Q}.
Then, for arbitrary single-qubit measurements,
it is included in FH${}_{3}$. 

Since $U_T \cdots U_1$ contains CNOT gates,
this readily tells us that 
an arbitrary CSS (Calderbank-Shor-Steane) state~\cite{SteaneCSS,CSS}
can be prepared from the final state via 
the measurement on the clock.
More generally, we have
\begin{theo}
An arbitrary state that is generated 
by a polynomial number of $X$, CNOT, and Toffoli gates
from $|0\rangle ^{\otimes n} |+\rangle ^{\otimes m}$
can be prepared efficiently by stoqAQC.
\end{theo}
If we are allowed to perform arbitrary single-qubit measurements,
we can show that universality with adaptive measurements 
using measurement-based quantum computation~\cite{MBQC}
and quantum computational supremacy with non-adaptive measurements~\cite{IQP0,IQP}
as follows.

\section{Universality with adaptive measurements}
\begin{figure}
\center{\includegraphics[width=0.8\linewidth]{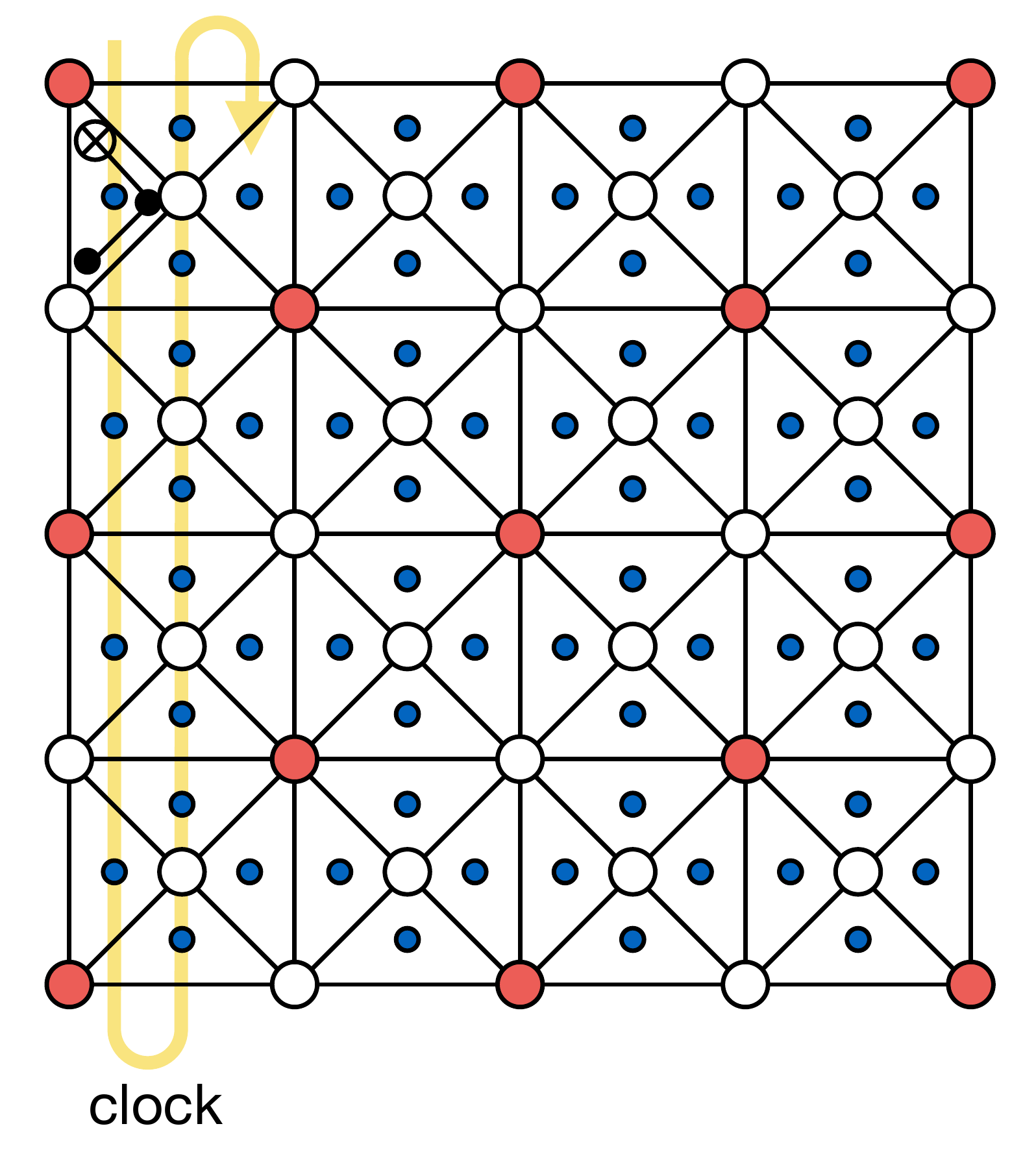}}%
\caption{A 6-local stoquastic Hamiltonian model, from which 
the union jack state is obtained from the final state. 
Since the union jack state is generated by a constant-depth 
circuit consisting of Toffoli gates, 
the domain wall clock qubits are intrinsically localized on the 
two-dimensional lattice.
\label{fig2}}
\end{figure}
Let us consider a more elaborate and concrete 
model with a 6-local Hamiltonian.
Let us consider a union jack lattice
as shown in Fig.~\ref{fig2},
where the qubits in the working and clock systems are located on 
the vertices (white and red circles) and face centers (blue circles)
of the triangles, respectively.
The qubits on the working system are prepared 
$|0\rangle$ and $|+\rangle$ on the white and red colored qubits,
respectively.
By introducing the domain wall clock
$|t\rangle = |1_1 \cdots 1_t 0_{t+1} \cdots \rangle ^{c}$
localized on the triangle centers~\cite{OliveiraTerhal},
the Hamiltonian is given by
\begin{eqnarray}
\tilde H_{\rm initial} &=& 
 H_{\rm initial} + H_{\rm clock}  
\\
\tilde H_{\rm final} &=&
H_{\rm final} + H_{\rm clock}
\end{eqnarray}
with the energy penalty terms for the illegal clock states,
\begin{eqnarray}
H_{\rm clock} = \sum _{i=1}^{T-1} |01\rangle \langle 01|^{c}_{i,i+1}.
\end{eqnarray}
The superscript $c$ denotes the clock system.
Moreover, following replacements in $H_{\rm initial}$ and $H_{\rm final}$ 
are also performed: 
\begin{eqnarray}
|0 \rangle \langle 0|_c 
& \rightarrow & |0\rangle \langle 0|^{c}_{1}
\\
I_c -|0 \rangle \langle 0|_c  & \rightarrow & |1\rangle \langle 1|^c_{1}
\\
|t\rangle \langle t-1|_c &\rightarrow& |110\rangle \langle 100 |^{c}_{t-1,t,t+1},
\\
|t-1\rangle \langle t-1 |_c &\rightarrow& |100 \rangle \langle 100| _{t-1,t,t+1}^{c},
\\
|t\rangle \langle t |_c &\rightarrow& |110 \rangle \langle 110| _{t-1,t,t+1}^{c},
\end{eqnarray}
where $|1\rangle _{t-1}$ and $|0\rangle _{t+1}$ 
are not required for the initial $t=0$ and final $t=T$ clock states, respectively.
Note that if a high energy penalty term for illegal clock states 
is introduced, the clock operator can be described by fewer body operators~\cite{KempeQMA}. However, the Hamiltonian can be finally
mapped into 2-local one perturbatively by using mediator qubits~\cite{stoqBravyi},
and hence we employ the original 5-local construction in Refs.~\cite{KitaevBook,AharonovAQC,OliveiraTerhal}.
$U_t$ in $H_{\rm final}$ 
corresponds to the Toffoli gate
acting
on three qubits on the triangle corresponding 
to the $t$-th clock qubit,
where white and red qubits act as the controls and target,
respectively, 
as shown in Fig.~\ref{fig2}.
Under the condition of projecting the clock state
to $|11...1\rangle^{c}$ on the final state,
the union jack state is obtained.
The union jack state is known to be a universal resource 
for measurement-based quantum computation with the Pauli basis measurements~\cite{MiyakeUJ}. 
Therefore, stoqAQC with adaptive Pauli measurements 
is universal. 
\begin{theo}
6-local StoqAQC with adaptive Pauli basis measurements can 
simulate universal quantum computation efficiently.
\end{theo}
The 6-local interactions can be reduced to 2-local 
perturbatively by using the mediator qubits~\cite{stoqBravyi}.

\section{Quantum computational supremacy with non-adaptive measurements}
The advantage of AQC 
would its protection against decoherence 
by the Hamiltonian.
However, the adaptive measurements 
considered above would deteriorate this good property,
since the resource state might decohere during the 
measurements. (Later, we will also see
how to make stoqAQC robust against the measurement imperfections.) 
If we consider a robust physical 
implementation, non-adaptive measurements on the 
final state would be preferred.
Even in such a case, 
we can show strong evidence of quantum speedup of stoqAQC
as follows. 

Measurement-based quantum computation on the union jack lattice and 
HC1Q are both known to be universal under postselection~\cite{MiyakeUJ,HC1Q}.
Therefore, non-adaptive measurements on the exact ground state $|\Psi\rangle$
are as powerful as postBQP~\cite{postBQP} under postseleciton.
This indicates that 
classical (non-adaptive) sampling in the Pauli bases
with a constant $l_1$ additive error
is impossible 
under complexity theoretical conjectures~\cite{IQP}:
\begin{theo}
Based on anti-concentration conjecture 
and average v.s. worst case conjecture
of HC1Q or IQP with the union jack state, 
efficient classical simulation of 
6-local stoqAQC with non-adaptive Pauli basis measurements 
with a small constant $l_1$ additive error 
implies the collapse of polynomial hierarchy to the third level.
\end{theo}
Again, the 6-local interactions can be reduced perturbatively to 2-local~\cite{stoqBravyi}.

\noindent{\it Proof.} 
Below we will first show that 
the sampling on the polynomial-time stoqAQC 
is sufficiently close to the output of the ideal circuit 
shown in Fig.~\ref{fig1} (a) in $l_1$-norm.
Then, a classical sampling of such stoqAQC
with a constant $l_1$ additive error implies 
a classical simulation of the output of the ideal circuit 
with a constant $l_1$ additive error.

Due to the adiabatic theorem, 
the final state $|\Psi _{\rm ad}\rangle$
satisfying 
\begin{eqnarray}
\| |\Psi _{\rm ad}\rangle  - | \Psi \rangle\|_2 <\epsilon 
\end{eqnarray}
can be obtained in polynomial time in the size of computation.
From this we have 
\begin{eqnarray}
|\langle \Psi_{\rm ad} | \Psi \rangle |> 1- \epsilon /2 \equiv f.
\end{eqnarray}
By using the relation between 
fidelity and trace distance,
we have
\begin{eqnarray}
\frac{1}{2} \| \rho  -  \tilde \rho  \|_1 = \sqrt{1- f^2 }\equiv \delta ',
\end{eqnarray}
where $\rho \equiv |\Psi \rangle \langle \Psi |$ and $\tilde \rho  \equiv
|\Psi _{\rm ad} \rangle \langle \Psi _{\rm ad} |$ are both pure states.
Let $P(x)=W^{\dag}|x\rangle\langle x|W$ be
projectors corresponding to the measurement outcome $x\in \{ 0,1\}^{n+m}$ 
on the working system
with $W$ being a product of single-qubit unitary operators for the basis change,
and $P_{\rm clock}$ be the projector corresponding to the successful clock states.
The ideal probability distribution of 
the circuit in Fig.~\ref{fig1} (a)
is given by 
\begin{eqnarray}
p(x) = {\rm Tr}[P(x)P_{\rm clock} \rho]/ p_{\rm clock},
\end{eqnarray}
where $p_{\rm clock} = {\rm Tr}[P_{\rm clock} \rho ]$.
Similarly, $\tilde p(x)$ and $\tilde p_{\rm clock}$
are defined for the final state $\tilde \rho $ 
of the adiabatic operation.
Since for any POVM (positive operator valued measure) operators $\{ M_y \}$,
we have 
\begin{eqnarray}
 \| q(y) - \tilde q (y) \|_1 
 \leq \| \rho - \tilde \rho \|_1
\end{eqnarray}
where $q(y) = {\rm Tr}[M_y \rho]$ and $\tilde q(y) = {\rm Tr}[M_y \tilde \rho]$.
Then we have 
\begin{eqnarray}
 \sum_{x} | p(x)p_{\rm clock} - \tilde p(x) \tilde p_{\rm clock}|
\leq  2\delta',
\label{eq02}
\end{eqnarray}
and 
\begin{eqnarray}
|p_{\rm clock} - \tilde p_{\rm clock}| \leq 2\delta ' .
\end{eqnarray}
On the other hand,
\begin{eqnarray}
&&\sum_x | p(x)p_{\rm clock} - \tilde p(x) \tilde p_{\rm clock}|
\nonumber \\
&\geq&\sum _x| [p(x) - \tilde p(x)]| p_{\rm clock} 
- |p_{\rm clock}- \tilde p_{\rm clock}|
\nonumber \\
&\geq& \sum _x |p(x) - \tilde p(x)] |  p_{\rm clock}
- 2\delta '.
\label{eq01}
\end{eqnarray}
By combining Eqs. (\ref{eq02}) and (\ref{eq01}),
\begin{eqnarray}
\| p(x) - \tilde p(x)\|_1 \leq 4 \delta' /p_{\rm clock}
\equiv \delta ''.
\end{eqnarray}
Therefore,
to achieve a polynomially small 
$l_1$ additive error $\delta '' = 1/{\rm poly}(n,m)$,
we need $\delta'< \delta '' p_{\rm clock}/4 $.
This means that the final state of an accuracy 
\begin{eqnarray}
\epsilon < 2 ( 1- \sqrt{1- (\delta ''p_{\rm clock}/4)^2})
\label{eq:accuracy}
\end{eqnarray}
in $l_2$-norm 
is enough.
This and Eq. (\ref{eq:adiabatic}) guarantees that 
the computation time of the stoqAQC is 
still polynomial time in $(n,m)$.

Suppose classical efficient sampling of
stoqAQC with a $l_1$ additive error $\eta$ is possible,
a similar argument tells us 
the conditional probability distribution 
$p_{\rm samp}(x)$ on the working system satisfies 
\begin{eqnarray} 
\| p_{\rm samp}(x) - \tilde p(x)\|_1 &<& 2\eta /\tilde p_{\rm clock} 
\\
&<&  2\eta /(p_{\rm clock} - 2 \delta' )  
\end{eqnarray}
where we assume $2\delta' < p_{\rm clock}$.
Note that $2 \delta ' = \delta '' p_{\rm clock} /2 $ is small enough.
Moreover, $p_{\rm clock}$ can be a constant,
and we chose $p_{\rm clock} = 1/2$.
Then, the $l_1$ additive error between 
conditional probability distributions 
from classical sampling $p_{\rm samp}(x)$ and the ideal one $p(x)$
is bounded by a constant value
\begin{eqnarray}
\| p_{\rm samp}(x) - p(x)\|_1
&<& \| p_{\rm samp}(x) - \tilde p(x)\|_1 + \| \tilde p(x) - p(x)\|_1
\nonumber\\
&<&  4\eta /(1 - 4 \delta' )  + \delta ''.
\end{eqnarray}
Both $\delta '$ and $\delta ''$
can be made small by improving the accuracy $\epsilon$ of the 
stoqAQC as seen in Eq.(\ref{eq:accuracy}), 
which is the target of the classical simulation.
Therefore, 
classical conditional sampling $p_{\rm samp}(x)$
is constantly close to $p(x)$ with $l_1$ additive error.
The probability distribution $p(x)$ 
contains IQP on the union jack lattice~\cite{MiyakeUJ} and HC1Q model~\cite{HC1Q},
both of which are shown to be postBQP-complete under postselection.
Therefore, 
by assuming that the anti-concentration and average v.s. worst case conjectures
are correct in these models,
a classical efficient sampling of stoqAQC 
with a small constant $l_1$ additive error 
leads to the collapse of the polynomial hierarchy to the third level.
\hfill $\blacksquare$

Here we should note that 
while a decision problem on stoqAQC 
with a single-qubit $X$-basis measurement on the working system
and computational basis measurements on the clock system
corresponds to stoqMA 
with a trivial proof $|0\rangle ^{\otimes n} |+\rangle ^{\otimes m}$,
and hence is upper bounded by postBPP~\cite{AQCreview,FarhiHarlow}.
Although this has been thought to be a partial evidence of 
the weakness of computational power of stoqAQC,
the sampling problem on stoqAQC with non-standard bases 
can be much harder leading to postBQP under postselection and 
exhibits quantum computation supremacy as seen above.

\section{Quantum speedup with non-adaptive measurements}
\begin{figure}[t]
\center{\includegraphics[width=1\linewidth]{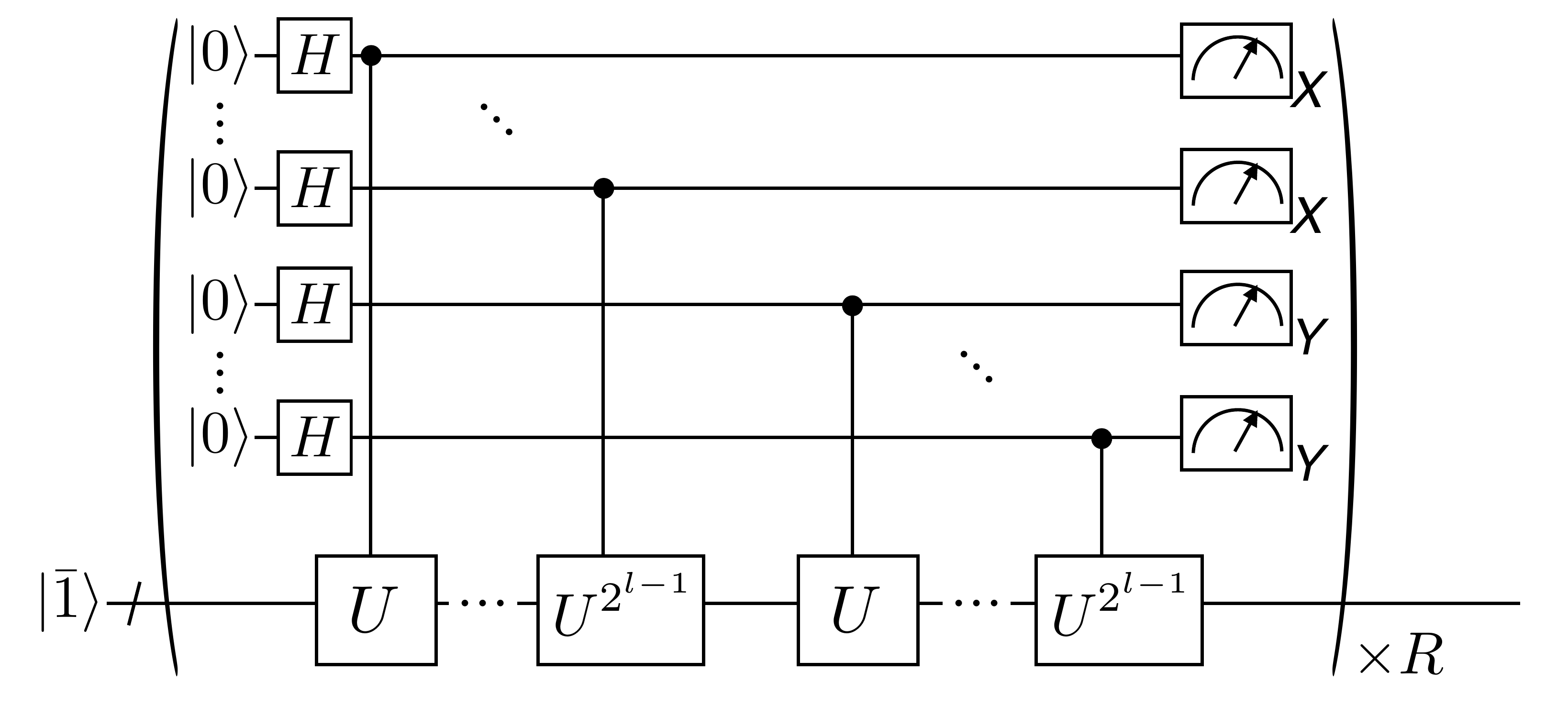}}%
\caption{Non-adaptive iterative phase estimation without quantum Fourier transformation.
\label{fig3}}
\end{figure}
Next we show that 
stoqAQC with non-adaptive single-qubit measurements 
can solve the factoring problem by simulating 
Shor's algorithm.
Since any classical reversible (unitary) computation can be 
done on the circuit shown in Fig.~\ref{fig1}, 
the modular exponentiation can also be implemented.
We employ the Kitaev's original 
phase estimation~\cite{KitaevPhase},
where adaptive single-qubit measurements 
are performed without the 
inverse quantum Fourier transformation.
Furthermore, 
in order to avoid the adaptive measurements, 
we introduce a non-adaptive iterative phase estimation 
as shown in Fig.~\ref{fig3}.

Let $N$ and $x$ be an integer to be factorized and 
its coprime respectively, 
and define $U_x = \sum _{y} |xy \textrm{ (mod $N$)} \rangle \langle y|$,
where $y=1,...,N$.
Let $r$ be the order, the minimum integer satisfying $x^r \equiv 1$
modulo $N$.
An eigenstate of $U_x$ with a label $s$ ($0\leq s \leq r-1$) is given by
\begin{eqnarray}
|u_s\rangle = \frac{1}{\sqrt{r}}\sum _{k=0}^{r-1} e^{- 2 \pi i (s/r)k} |x^k \textrm{(mod $N$)}\rangle.
\end{eqnarray} 
The eigenvalue $e^{2 \pi i (s/r)}$ is 
written in terms of the binary representation 
by $e^{(2 \pi i )0.j^{(s)}_1 ... j^{(s)}_l }$ ($j^{(s)}_k \in \{ 0,1\}$).
As usual, the phase estimation is done on the state
\begin{eqnarray} 
|\bar 1 \rangle = \frac{1}{\sqrt{r}}\sum _{s=0}^{r-1} |u_s\rangle,
\end{eqnarray}
where the notation $|\bar 1\rangle$
is employed to avoid the confusion with the basis state $|1\rangle$ of a qubit.

As shown in Fig.~\ref{fig3},
we perform $2R$ iterative phase estimation 
of $U^{2^k}$
with non-adaptive measurements for each $k=0,...,l-1$.
The first and second $R$ iterations are measured 
in the $X$ and $Y$ basis non-adaptively.
The state before the measurement is written by 
\begin{widetext}
\begin{eqnarray}
\frac{1}{\sqrt{r}}\sum _{s=0}^{r-1}
\left(\frac{|0\rangle + e^{(2\pi i) 0.j^{(s)}_1 ... j^{(s)}_l} |1\rangle}{\sqrt{2}}
\right)^{\otimes 2R} 
\cdots 
\left(\frac{|0\rangle + e^{(2\pi i) 0.j^{(s)}_k ... j^{(s)}_l} |1\rangle}{\sqrt{2}}
\right)^{\otimes 2R} 
\cdots
\left(\frac{|0\rangle + e^{(2\pi i) 0. j^{(s)}_l} |1\rangle}{\sqrt{2}}
\right)^{\otimes 2R} | u_s\rangle.
\end{eqnarray}
\end{widetext}
While the measurements are done non-adaptively,
we below employ an inductive argument for explanation.
Let us first consider the measurements on 
\begin{eqnarray}
\left(\frac{|0\rangle + e^{(2\pi i) 0. j^{(s)}_l} |1\rangle}{\sqrt{2}}
\right)^{\otimes 2R}.
\end{eqnarray}
Since $j^{(s)}_l$ is either $0$ or $1$,
we can easily determine $j^{(s)}_l$ 
from the $R$ measurement outcomes in each $X$ and $Y$ bases.
The $R$ samples are enough to decide $j^{(s)}_l=0,1$ with an 
exponential accuracy by using Hoeffding's inequality,
since expectation values of the angle
$\theta _l \equiv \arctan (\langle Y\rangle / \langle X \rangle)$ 
is constantly separated by $\pi$ for $j^{(s)}_l=0$ or $=1$.
We can think this process just as a tomography of
\begin{eqnarray}
\frac{|0\rangle + e^{(2\pi i) 0. j^{(s)}_l } |1\rangle}{\sqrt{2}}
\end{eqnarray}
using $2R$ samples on the $xy$-plane of the Bloch sphere.

Next, suppose $j^{(s)}_{k-1},...,j^{(s)}_l$ are all determined already.
Then we perform the sampling on 
\begin{eqnarray}
\left(\frac{|0\rangle + e^{(2\pi i) 0.j^{(s)}_k ... j^{(s)}_l} |1\rangle}{\sqrt{2}}
\right)^{\otimes 2R} .
\end{eqnarray}
Since $j^{(s)}_{k-1},...,j^{(s)}_l$ are determined already,
the expectation value of the angle
\begin{eqnarray}
\theta _{k} = \arctan (\langle Y \rangle / \langle X \rangle ),
\end{eqnarray}
evaluated by the state
\begin{eqnarray}
\left(\frac{|0\rangle + e^{(2\pi i) 0.j^{(s)}_k ... j^{(s)}_l} |1\rangle}{\sqrt{2}}
\right),
\end{eqnarray}
is again constantly separated by $\pi$ for $j^{(s)}_k=0,1$ cases.
Therefore, from $2R$ measurement outcomes,
we can determine $j^{(s)}_k$ with an exponential accuracy 
in the number of samples $R$ by using Hoeffding inequality 
for $\langle X\rangle$ and $\langle Y \rangle$.
Inductively, all binary bits, $j_1^{(s)},...,j_{l}^{(s)}$, are determined.
Again, we should note that the measurements are non-adaptive
and hence can be done simultaneously.
From the obtained phase,
the order $r$ is obtained by using the continued fraction as usual.
Then, a non-trivial factor of $N$ is found with a high probability
against a random choice of $x$.

The circuit shown in Fig.~\ref{fig3} 
consists only of $X$, CNOT, and Toffoli
acting on the initial $|0\rangle$s and $|+\rangle$s.
Therefore, including the measurements on the clock,
we conclude that 
stoqAQC with non-adaptive single-qubit measurements 
can solve the factoring problem in polynomial time 
by simulating Shor's algorithm.
If measurement bases are restricted to $X$ and $Z$,
still stoqAQC with non-adaptive measurements 
can solve Simon's problem~\cite{Simon}.
Since swap operations can be constructed from CNOT gates,
the stoquastic Hamiltonian can be reduced to spatially two-local one on 
a two-dimensional lattice by localizing the domain wall clock~\cite{OliveiraTerhal}
and applying perturbative approach with the mediator qubits~\cite{stoqBravyi}.

\section{Verification and measurement error tolerance}
Our construction
based on the Feynman-Kitaev construction
naturally provides a verification 
protocol of whether or not the final state of stoqAQC is 
faithfully generated.
Several verification protocols 
with single-qubit measurements 
have been proposed so far 
based on the Kitaev-Feynman construction~\cite{PostHoc,MorimaeTakeuchi}
and can be readily applied to the present stoqAQC Hamiltonians.

The history state $|\Psi \rangle$
is the unique ground state of the final Hamiltonian $H_{\rm final}$
and satisfies 
\begin{eqnarray}
E_g= \langle \Psi | H_{\rm final} |\Psi \rangle = 0.
\label{eq:enegy}
\end{eqnarray}
Any pure state can be 
expanded by the energy eigenstates $\{ |E_i \rangle \}$
with eigenvalue $\{E_i\}$, respectively:
\begin{eqnarray}
|\psi \rangle = \alpha _0 |E_0 \rangle + \sum _{i=1}^{(T+1)2^{(n+m)}-1} 
\alpha _i | E_i \rangle,
\end{eqnarray}
where $|E_0\rangle = |\Psi \rangle$ and $E_0 =0$.
The energy expectation value becomes 
\begin{eqnarray}
\sum _{i=0}^{(T+1)2^{(n+m)}-1}  |\alpha _i |^2 E_i > (1-|\alpha_0|^2) \Delta .
\end{eqnarray}
(Its extension to the mixed states is straightforward.)
Therefore, by measuring the energy expectation value $E_{\rm exp}$ 
of the final state
within an additive error 
and by checking whether or not 
\begin{eqnarray}
E_{\rm exp} < \epsilon_{\rm ver} \Delta
\end{eqnarray}
is satisfied with polynomially small $\epsilon_{\rm ver}= 1/{\rm poly(n,m)}$,
we can verify that the overlap between 
the experimentally obtained state $|\psi\rangle$
and the ideal final state $|\Psi \rangle$ is 
sufficiently large $|\alpha _0|^2 > 1-\epsilon_{\rm ver}$.
Note that the fidelity is given by 
$|\langle \psi | \Psi \rangle|=|\alpha _0|$.
The measurement of the energy can be 
done by polynomially repetitive single-qubit measurements on the final state
as done in variational quantum eigensolver~\cite{VQE,GVQE}.
A more elaborated verification scheme as done in Ref.~\cite{MorimaeTakeuchi}
can also be employed.

Furthermore, 
as mentioned before,
if the measurement basis is restricted 
to the $xy$-plane or the $Z$ basis,
stoqAQC belongs to FH$_2$.
Therefore, the output $x$ 
can be verified efficiently 
by using the fact that a decision problem in FH$_2$ is in MA~\cite{VerificationFH,HC1Q}.

Finally, we consider the robustness of the proposed 
stoqAQC model against a measurement imperfection.
Even if AQC is executed ideally,
the final measurements, especially
done in the non-standard Pauli bases, would causes 
an imperfect readout.
Fortunately,
the measurements are done all in the Pauli bases.
The encoding circuit of CSS codes, such as 
the Steane 7-qubit code~\cite{SteaneCSS},
consists only of CNOT gates acting on 
$|0\rangle$s and $|+\rangle$s.
Therefore,
by adding the further working and clock qubits,
we can encode each qubit in the final state into 
a self-dual CSS code as shown in Fig.~\ref{fig1} (b).
For the self-dual CSS code, like the Steane 7-qubit code,
all logical Pauli basis measurements are done 
transversally by single-qubit Pauli basis measurements.
Therefore quantum speedup of stoqAQC is at least robust against 
the measurement imperfections in the non-standard basis.
While we employ non-fault-tolerant encoding circuit of the CSS code,
the encoded state is prepared as a ground state in AQC,
and hence it would be interesting to see whether or not 
this type of encoding improves the accuracy of computation 
against imperfections during adiabatic operations or 
finite temperature effects.

\section{Discussion}
Here we characterized computational power of 
stoqAQC with adaptive or non-adaptive 
single-qubit measurements in the non-standard bases.
While our stoqAQC model employs 
the 6-local stoquastic Hamiltonian,
it is straightforward to reduce it to 
a 2-local stoquastic Hamiltonian as shown in Ref.~\cite{stoqBravyi}.
While the mediator qubits are added,
the original Hamiltonian is simulated perturbatively 
on the original system. 
Hence single-qubit measurements are enough for our purpose.
However, in the case of the reduction from the 2-local stoquastic Hamiltonian 
to a transverse Ising model (on degree-3 graphs) shown in Ref.~\cite{stoqBravyi17},
each qubit is simulated by dual rail bosons, a hard-core dimer
and an Ising spin (chain). 
This modifies the single-qubit measurements 
in the original model
to non-local measurements on the mapped models.
Therefore, 
the computational power of stoqAQC with 
single-qubit measurements in non-standard bases
is open for the transverse Ising models.

Our result implies that 
non-stoquasticity is not necessarily required 
to get quantum speedup in an AQC machine.
Instead, quantum coherence of the final state and 
non-standard basis measurements are key ingredients.
Since the lowest energy gap closes inverse polynomially, 
a finite temperature effect would be crucial.
A fault-tolerance theory would be further required to 
achieve scalability.
An arbitrary CSS state can be prepared in an adiabatic way,
which would be interesting itself as 
a robust entangled state generation scheme
using only two-body stoquastic interactions.
Moreover, 
our construction can also be viewed as 
measurement-based quantum computation 
on a ground state of two-body Hamiltonians
~\cite{Nest08,Griffin08,Chen09,MiyakeAKLT,Cai10}.
In such a context, 
thermal equilibrium states are also known 
as universal resources~\cite{Li11,Fujii12,FujiiNakata}.
These constructions might be helpful 
in a construction of a fault-tolerant theory
of stoqAQC at finite temperature.
In addition, the embedding of the CSS state would be also 
useful for this purpose.

Finally, it would be interesting to see
whether or not quantum Monte-Carlo or other classical methods 
relevant for stoquastic Hamiltonians can cope with
the final basis change at the measurements.
It would be natural to conjecture that 
stoqAQC with a polynomially bounded lowest energy gap
can be classified into
three classes by the types of the measurements:
(i) classically simulatable with standard basis measurements, 
(ii) non-universal but quantum computational supremacy with non-adaptive non-standard basis measurements,
and (iii) universal with adaptive non-standard measurements.

In the most models exhibiting quantum computational supremacy~\cite{boson,IQP,FujiiTamate,FujiiPost,DuanIQP,Boixo,AaronsonRC,BoulandRC},
the sampled output itself is not so useful, 
except for the one-clean qubit model~\cite{DQC1,DQC12,DQC1hardness}
and HC1Q~\cite{HC1Q}.
The present result has pushed 
stoqAQC to one of the most powerful intermediate model of quantum computation,
which exhibits quantum computational supremacy and 
can solve a meaningful problem, such as the factoring problem.

\begin{acknowledgments}
The author would like to thank
Keiji Matsumoto, Tomoyuki Morimae, Yuki Takeuchi, and Shuhei Tamate
for valuable discussions and comments on the draft.
This work is supported by KAKENHI No. 16H02211, JST PRESTO JPMJPR1668, JST ERATO JPM- JER1601, and JST CREST JPMJCR1673.
\end{acknowledgments}

\bibliographystyle{apsrev4-1}
\bibliography{stoqAQC}

\end{document}